\documentclass[english]{svjour3}
\usepackage[T1]{fontenc}
\usepackage[latin9]{inputenc}
\usepackage{color}
\usepackage{verbatim}
\usepackage{prettyref}
\usepackage{mathrsfs}
\usepackage{url}
\usepackage{amsmath}
\usepackage{amssymb}
\PassOptionsToPackage{normalem}{ulem}
\usepackage{ulem}

\makeatletter
\usepackage{lineno}

\usepackage{esint}

\makeatother

\usepackage{babel}
\begin{document}

\title{\noindent Bohmian trajectories for Kerr-Newman particles in complex
space-time}

\author{\noindent Mark Davidson}

\institute{\noindent Mark Davidson\at Spectel Research Corp., Palo Alto, CA,
USA\\
\email{mdavid@spectelresearch.com}\\
}

\date{\noindent Received: date / Accepted: date}
\maketitle
\begin{abstract}
\noindent %

\noindent \textcolor{black}{Complexified Li\'{e}nard\textendash Wiechert
potentials simplify the mathematics of Kerr-Newman particles. Here
we constrain them by fiat to move along Bohmian trajectories to see
if anything interesting occurs, as their equations of motion are not
known. A covariant theory due to Stueckelberg is used.} This paper
deviates from the traditional Bohmian interpretation of quantum mechanics
since the electromagnetic interactions of Kerr-Newman particles are
dictated by general relativity. A Gaussian wave function is used to
produce the Bohmian trajectories, which are found to be multi-valued.
A generalized analytic continuation (GAN) is introduced which leads
to an infinite number of trajectories. These include the entire set
of Bohmian trajectories. This leads to multiple retarded times which
come into play in complex space-time. If one weights these trajectories
by their natural Bohmian weighting factors, then it is found that
the particles do not radiate, that they are extended, and that they
can have a finite electrostatic self energy, thus avoiding the usual
divergence of the charged point particle. This effort does not in
any way criticize or downplay the traditional Bohmian interpretation
which does not assume the standard electromagnetic coupling to charged
particles, but it suggests that a hybridization of Kerr-Newman particle
theory with Bohmian mechanics might lead to interesting new physics,
and maybe even the possibility of emergent quantum mechanics.

\noindent  
\end{abstract}

\section{Introduction\label{sec:Introduction}}

\noindent This paper studies Kerr-Newman charged particles moving
along free-particle Bohmian trajectories embedded in complex space-time.
These are coupled directly to the electromagnetic fields in the standard
way, and the resulting Liénard-Wiechert potentials are analyzed. This
form of coupling is dictated by the theory of general relativity.
In the Bohmian interpretation of quantum mechanics, no such coupling
of Bohmian particles to classical electromagnetic fields is assumed.
Therefore, in this paper, I am deviating from the Bohmian interpretation,
but using the Bohmian trajectories nonetheless. The Bohmian interpretation
is perfectly consistent, and I do not mean by this to detract from
it in any way. Rather I'm trying to see if there is anything to be
learned about Kerr-Newman particles in this way. It appears that there
might be. 

Recently Maldacena and Susskind have proposed that quantum entanglement
and the Einstein-Rosen bridge might be intimately related to one another
\cite{maldacena_cool_2013,susskind_dear_2017}. They and others argue
that gravity may be an emergent phenomenon, and that quantum entanglement
is a crucial ingredient of this emergence \cite{raamsdonk_building_2010,cowen_quantum_2015,verlinde_emergent_2017}.
Einstein thought that quantum theory might possibly be derivable from
general relativity and electromagnetism. This is clear from some of
the statements in the Einstein and Rosen paper \cite{einstein_particle_1935},
as well as many other writings of Einstein dealing with unified field
theory \cite{sauer_einsteins_2014}. The quest for a deeper understanding
of quantum mechanics is still an active research topic. I cite only
a small but significant sampling along these lines \cite{t_hooft_cellular_2014,rovelli_relational_1996,weinberg_collapse_2012,penrose_road_2007,penrose_fashion_2016}.
The results presented here touch upon these issues, and on the possibility
of emergent quantum mechanics.

This paper is a follow-up to an earlier study \cite{davidson_lorentz-dirac_2012,davidson_study_2012}
in which the complexified Lorentz-Dirac equation was analyzed, and
by means of analytic continuation some of the runaway solutions became
oscillatory, providing a mechanism for zitterbewegung, whose importance
for the foundations of quantum mechanics has been emphasized \cite{hestenes_zitterbewegung_1990,derakhshani_suggested_2015}.
The analytic continuation was justified by complex manifold techniques
that have been prevalent in the Kerr-Newman metric literature \cite{adamo_null_2009,newman_heaven_1976,newman_complex_1973}.
In this paper, I extend these results by incorporating Bohmian trajectories
for the Kerr-Newman particles. The idea that elementary particles
might be Kerr-Newman singularities in general relativity has been
boldly, eloquently, and persistently championed by Alexander Burinskii
\cite{burinskii_microgeon_1974,burinskii_kerr_1998,burinskii_dirac-kerr_2005,burinskii_dirac_2008,burinskii_gravitational_2011,burinskii_regularized_2010}.
He also believes strongly in the connection between string theory
and his Kerr-Newman electron theory, since the locus of singularities
form one dimensional rings. The idea proposed here is that these particles
may be traveling along Bohmian trajectories moving in complex space-time.
Reviews of Bohmian mechanics are given in \cite{bacciagaluppi_quantum_2009,bohm_undivided_1995,durr_quantum_2012,holland_quantum_1995}.
I want to develop a covariant theory here, and the original Bohm model,
being non-relativistic, is not suitable. Although there are Bohmian
models for the Dirac and Klein-Gordon equations in the literature,
I have chosen for this first attempt a trajectory theory based on
the relativistically covariant ``proper time'' wave mechanics of
Fock and Stueckelberg \cite{fock_eigenzeit_1937,stueckelberg_signification_1941}.
An excellent review is the book by Horwitz \cite{horwitz_relativistic_2015}.
The application of Bohmian mechanics to this system was done by Kyprianidis
and Fanchi in \cite{kyprianidis_particle_1985,kyprianidis_scalar_1987,fanchi_quantum_2000}.
Holland has made some critical comments in \cite{holland_quantum_1995},
but I choose to work with it despite these, because the Stueckelberg
approach allows separability in all four space-time axes, and this
facilitates finding exact solutions needed for analytic continuation
without errors. The problem of relativistic Bohmian mechanics was
also considered in \cite{durr_can_2013}, and other approaches proposed
there. Those could be considered in the future. In all of these Bohmian
references there is no assumed coupling between the charged particles
and the classical electromagnetic fields. In this paper I am borrowing
the trajectories from the Bohmian interpretation, but then I'm deviating
from the Bohm interpretation by treating the particles as standard
Kerr-Newman particles in complex space-time which do couple to the
classical electromagnetic fields. My initial interest in doing this
was curiosity, but I would say that now the results obtained in this
way are interesting. But it should not be taken as implying that the
standard Bohmian interpretation of quantum mechanics should be modified
by this, because it doesn't need such modification.

The null retarded and advanced time solutions determine the Liénard-Wiechert
potentials in classical electromagnetism. In real Minkowski space,
there are generally only two null solutions - one advanced and one
retarded - for a given observation point and for a single classical
timelike particle. In the complexified space-time there can be many
complex-valued null solutions for a given trajectory, and the goal
here is to study these. Since such particles would be very light,
their gravitational distortion of the metric would be ignorable to
a good approximation. But, the lingering effect that cannot be ignored
is the complex space-time embedding. If there are multiple Riemann
sheets for the Liénard-Wiechert potentials in complex space-time,
as was found for example in \cite{davidson_lorentz-dirac_2012}, then
this feature cannot be ignored, even when the metric distortions due
to particle masses can be ignored. This fact has been largely overlooked
in the physics literature because multi-valued fields and trajectories
in complex space-time have not been given serious consideration previously\footnote{The static Kerr-Newman solution to the electromagnetic field is double
valued when the observation point is continued along paths in space-time.
This is well known and studied. Here we are talking about multi-valued
analytic functions of the world time variable when the trajectory
is a function of this time and not static.}. So, in this paper, I shall work in complex Minkowski space, and
consequently the mathematical description is tremendously simplified
compared with curved metrics. But the lingering and critically important
influence of gravity and general relativity is the complex space-time
embedding, and the multi-valued Riemann sheet structure of both particle
trajectories and their electromagnetic field interactions. These effects
are not proportional to the gravitational constant, and can therefore
affect measurements significantly in the small particle limit. Nevertheless,
they can still be interpreted as gravitational effects.

The reader may reasonably ask if I'm proposing here that the complex
space-time is actually an ontological reality. Personally, I'm agnostic
on that point. One of the principle adherents of complex space-time
approach, Ezra Newman, does not believe it's real, but that it's an
extremely useful construct for mathematically analyzing the equations
of general relativity for reasons that remain somewhat obscure \cite{newman_private_2017}.
In section \ref{sec:Interpretation-of-complex} I offer my own more
detailed thoughts on this issue.

Kerr-Newman solutions are obtained most easily by considering point
particles in complex Minkowski space. In all cases covered by the
Kerr-Newman solutions (i.e. all values of mass, charge, and spin that
define a solution), the particle is first considered as a structureless
point particle. Spin arises by displacing the particle from the real
axis. This is why I have chosen to begin here with the Klein-Gordon-Stueckelberg
equation rather than a spin 1/2 version, since spin implies some sort
of internal structure to the particle, but the Klein-Gordon solution
does not. It seemed like a good place to start. My ultimate goal is
to try and describe the spin 1/2 case as well, and I am studying this
problem, but am not sure precisely how to proceed. I'm interested
in possibly combining the results of \cite{davidson_lorentz-dirac_2012},
which gave a model for zitterbewegung within the Kerr-Newman framework,
together with Bohmian trajectories. Alexander Burinskii has a different
model for zitterbewegung which is based on a gravitating bag model
involving additional field effects \cite{burinskii_gravitating_2015,burinskii_emergence_2015},
and he has argued that this leads to an explanation for Dirac equation.
The bag structure gives internal structure to the particle, and it's
no longer a simple point in complex space-time. I've chosen to work
with the Stueckelberg formalism because it satisfies Wigner locality,
and is manifestly Lorentz covariant, and in addition the solutions
are simpler because there is no mass shell constraint, and this allows
separable solutions along all four space-time axes, without the mass
constraint messing up separability. I wanted to work with exact closed-form
solutions so that analyticity could be exactly studied, and analyticity
cannot be studied with approximate solutions. This is a starting point.
Future research can examine various ways to introduce spin. If this
approach proves fruitful, then it may lead to a second application
of Bohmian mechanics which links it to classical general relativity
in a way that might point to a path to emergent quantum mechanics.

\section{Covariant Bohmian mechanics}

\noindent Following \cite{kyprianidis_particle_1985,fanchi_quantum_2000},
we first develop a Hamilton-Jacobi description of the many-particle
covariant Stueckelberg equation in Minkowski space (metric signature
{[}+,-,-,-{]} with c=1). This version of relativistic mechanics, both
classical and quantum, goes by other names such as Fock-Stueckelberg,
proper time formalism, world or historical time formalism, Stueckelberg-Horwitz-Piron
theory, covariant Schrödinger equation, etc. A fine recent review
is given by Horwitz \cite{horwitz_relativistic_2015}. It introduces
an additional Lorentz invariant and universal world time, which is
denoted here by the variable $s$. It plays a role similar to the
proper time in relativistic classical mechanics, but it is not the
same as proper time. The basic wave equation is a generalization of
the many particle Schrödinger equation

\noindent 
\begin{equation}
i\hbar\frac{\partial\Psi}{\partial s}=K\Psi\label{eq:Fanchi equation}
\end{equation}

\date{\noindent 
\begin{equation}
K=\sum_{a=1}^{N}\frac{1}{2m_{a}}\left(-i\hbar\frac{\partial}{\partial x_{a\mu}}-\frac{e}{c}A_{a}^{\mu}\right)\left(-i\hbar\frac{\partial}{\partial x_{a}^{\mu}}-\frac{e}{c}A_{a\mu}\right)+U(x_{1},...,x_{N})
\end{equation}
where the momentum operator for particle $a$ is $p_{a}^{\mu}=-i\hbar\partial/\partial x_{a\mu}$.
We follow the sign convention and metric signature used in \cite{fanchi_quantum_2000}
with no loss of generality. }

The wave function $\Psi$ depends on a single Lorentz invariant ``world
time'' parameter $s$ as well as the N-tuple of Minkowski coordinates
$x_{a\mu}$. It is assumed to be a single valued function of the $x_{a\mu}$,
at least when they are restricted to real values. $A_{a}^{\mu}$ is
the external 4-vector potential acting on particle $a$. The extra
potential U describes an additional non-electromagnetic potential
term. All potentials are assumed to be analytic functions of all of
their arguments. The normalization condition is 

\noindent 
\begin{equation}
\int\Psi^{*}(x_{1},...,x_{N},s)\Psi(x_{1},...,x_{N},s)d^{4}x_{1}...d^{4}x_{N}=1
\end{equation}

\noindent In the standard Bohmian trajectory approach, followed in
\cite{fanchi_quantum_2000}, one writes

\noindent 
\begin{equation}
\Psi(x,s)=R(x,s)e^{iS_{B}(x,s)/\hbar},\:S_{B}=\hbar Im(ln(\Psi))\label{eq:Bohm Hamilton-Jacobi function}
\end{equation}

\noindent where for all $x_{a}$ and s real, $R$ and $S$ are real,
$\rho=R^{2}$, and $R\geq0$. The conservation equation is

\noindent 
\begin{equation}
\frac{\partial\rho}{\partial s}+\sum_{a=1}^{N}\partial_{a\mu}V_{a}^{\mu}\rho=0,\:V_{a}^{\mu}=\frac{1}{m_{a}}\left(\partial_{a}^{\mu}S_{B}-\frac{e}{c}A_{a}^{\mu}\right)
\end{equation}

\noindent The vector potentials $A_{a}^{\mu}$ as seen by particle
a will depend not only on $x_{a}$, but also in general on the other
$N-1$ coordinates of the other interacting particles. These equations
cannot be considered a complete description of the electromagnetic
interaction, since the Liénard-Wiechert potentials for inter-particle
interaction involve retarded (or maybe even advanced) times, and for
relativistic particles such times depend not only on the particle
positions but also on their entire trajectories for their calculation.
But in quantum theory we do not have a trajectory to work with. One
way to remedy this situation is to develop a second quantized field
theory to handle the interactions, as is done in QED. This has led
to a 5d version of electromagnetism called pre-Maxwell theory \cite{land_pre-maxwell_1998,land_greens_1991,land_offshell_2013}.
If we try to use the Bohmian trajectories in real space time to calculate
the electromagnetic fields produced, we quickly get unphysical results.
For example, for a single free quantum particle, the Bohmian trajectories
in general will be accelerated (and/or decelerated) by the quantum
mechanical potential, and if they are charged then they will radiate
electromagnetically and lose energy, but this clearly doesn't happen
for free particles as Ehrenfest's theorem proves that they don't slow
down in quantum mechanics, and experiments support this. In this paper,
we consider the fields created by Bohmian trajectories, not in real,
but in complex space-time where the radiation formulas are different.
It is my hope that this will lead to new and interesting results,
and perhaps allow the Bohm trajectories to be interpreted as classical
sources for the electromagnetic fields. The reason that complex space-time
makes a difference is that the Liénard-Wiechert potentials can have
more than the two null-time solutions which they have in real space
time. This comes about because of the possibility of multi-valued
retarded times and even multi-valued trajectories in the complex case.

\noindent We can create a single particle current density by integrating
the continuity equation over the other N-1 particle positions. Suppose
we want to integrate over all but particle $a$. We obtain

\noindent 
\begin{equation}
\rho_{a}(x_{a},s)=\int\left(\prod_{m\neq a}d^{4}x_{m}\right)\rho(x_{1},...,x_{N},s)
\end{equation}

\noindent 
\begin{equation}
J_{a}^{\mu}(x_{a},s)=\int\left(\prod_{m\neq a}d^{4}x_{m}\right)V_{a}^{\mu}(x_{1},...,x_{N},s)\rho(x_{1},...,x_{N},s)\label{eq:Bohmian current density}
\end{equation}
so the single particle continuity equation becomes

\noindent 
\begin{equation}
\frac{\partial\rho}{\partial s}+\partial_{a\mu}J_{a}^{\mu}(x_{a},s)=0,\:\partial_{a\mu}\int_{-\infty}^{+\infty}J_{a}^{\mu}(x_{a},s)ds=0\label{eq:Particle continuity equation}
\end{equation}

\noindent So we can identify the electromagnetic 4-current density
for particle a by

\begin{equation}
\mathcal{J_{\mathrm{\mathit{a\,}}\mathcal{EM}}^{\nu}}(x)=q_{a}\int_{-\infty}^{+\infty}J_{a}^{\mu}(x_{a},s)ds\label{eq:Charge Normalization I}
\end{equation}

\noindent and where $q_{a}$ is a constant which is just the charge
of the particle $Q_{a}$ up to a normalization factor.

\begin{equation}
Q_{a}=\int d^{3}x\mathcal{J_{\mathrm{\mathit{a\,}}\mathcal{EM}}^{\mathit{0}}}(x)\label{eq:Charge normalization II}
\end{equation}

\noindent The Bohmian trajectory equation is the solution to the differential
equation

\begin{equation}
\frac{dx_{a}(s)^{\mu}}{ds}=\frac{1}{m_{a}}\left(\frac{\partial}{\partial x_{a\mu}}S_{B}(x,s)-\frac{q_{a}}{c}A_{a}^{\mu}(x)\right)\label{eq:Bohmian velocity equation}
\end{equation}

\noindent One distinguishing characteristic of Fock-Stueckelberg wave
mechanics is that the rest masses of particles are not fixed, but
can change due to interactions. The off-mass-shell possibility in
quantum theory has been analyzed in a series of papers by Greenberger
\cite{greenberger_theory_1970-1,greenberger_theory_1970,greenberger_useful_1974,greenberger_wavepackets_1974}.
The classical electromagnetic interaction alone cannot change the
mass, but if the potential term $U(x_{1},...,x_{N})$ is weak, then
if the masses are initially on or near their mass shell values $m_{a}$,
then as the world time variable $s$ advances, the mass shell constraint
will remain approximately satisfied so that $m_{a}\approx p_{a}^{\mu}p_{a\mu}$,
as described in \cite{horwitz_relativistic_2015}. In this way the
Stueckelberg theory includes the the Klein-Gordon equation as a limiting
case when the masses can be taken as fixed. The advantage of the variable-mass
Stueckelberg equation as opposed to the Klien-Gordon or other fixed
mass equations is that Newton-Wigner locality holds for the position
operators \cite{horwitz_relativistic_2015} while maintaining manifest
covariance. It is also very similar to the Schrödinger equation, being
first order in $s$ derivatives, and so many of the results of the
usual Bohmian mechanics are easily carried over. The question of what
draws particles back to the mass shell if they have somehow wandered
off it is an area of active current research \cite{horwitz_statistical_2016,land_speeds_2016}.

These functions $\Psi,\:R,\:\textrm{and}\:S_{B}$ in \prettyref{eq:Bohm Hamilton-Jacobi function},
defined for real arguments, are assumed to be sufficiently analytic
locally to allow analytic continuation to all complex values for $x$
and $s$. This technique was used, for example, in \cite{bender_complex_2002,swanson_transition_2004,poirier_flux_2008,chou_complex-extended_2010}.
$R$ and $S$ can then become complex valued, and the resulting trajectories
will trace out a curve in real space-time for real values of s, and
for complex values of s, it will describe a 2D surface embedded in
the 8D complex Minkowski space. The projection of this 2D surface
onto the real hyperspace will look like a string, a point often emphasized
by Burinskii \cite{burinskii_kerr_1998}.

\noindent \textbf{The free particle case}

\noindent 
\begin{equation}
i\hbar\frac{\partial\Psi}{\partial s}=-\frac{\hbar^{2}}{2M}\partial_{\mu}\partial^{\mu}\Psi=-\frac{\hbar^{2}}{2M}\left[\partial_{0}^{2}-\partial_{1}^{2}-\partial_{2}^{2}-\partial_{3}^{2}\right]\Psi=\frac{\hat{p}^{\mu}\hat{p}_{\mu}}{2M}\Psi\label{eq:Free particle Stueckelberg}
\end{equation}
We can consider a basis set of plane wave solutions, as usual, except
that the mass is not constrained to be on the mass shell here

\noindent 
\begin{equation}
\Psi(x,s)=\int\frac{d^{4}p}{\left(2\pi\hbar\right)^{4}}\tilde{\Psi}(p,s)e^{ip^{\mu}x_{\mu}/\hbar}
\end{equation}
In general, $p^{\mu}p_{\mu}\neq M^{2}$. The operator for momentum
is simply $\hat{p}^{\mu}=-i\hbar\frac{\partial}{\partial x_{\mu}}$.
We have a straightforward uncertainty relation for each of the four
coordinates, including a time-energy relation 

\noindent 
\begin{equation}
\sigma_{x^{\mu}}\sigma_{p^{\mu}}\geq\frac{\hbar}{2},\:\mu\in\{0,1,2,3\}
\end{equation}

\noindent The rest mass associated with a plane wave solution is $m^{2}=p^{\mu}p_{\mu}$,
with no limitation on the four independent values $p^{\mu}$. Negative
energy particles are interpreted as moving backwards in time, which
are in turn interpreted as anti-particles, an idea originally due
to Stueckelberg which has long since been incorporated into quantum
field theory. Negative $m^{2}$ particles are considered as tachyons.
The question of how the particle mass avoids unobserved off-shell
behavior is an area of current research \cite{horwitz_statistical_2016,land_speeds_2016,aharonovich_radiation-reaction_2012}.
These plane wave states are eigenstates for the $\hat{m}^{2}=\hat{p}^{\mu}\hat{p}_{\mu}$
operator. We say that if a state is built up by linear superposition
from plane waves whose mass is given by $m^{2}=M^{2}$, then the particle
is ``on its mass shell''. For all eigenstates of $\hat{m}^{2}$,
the wave equation \prettyref{eq:Free particle Stueckelberg} becomes
the standard Klein-Gordon equation by separating variables

\noindent 
\begin{equation}
\varPsi(x,s)=\psi_{m}(x)e^{-ism^{2}/2M\hbar},\:\left[m^{2}-\hat{p}^{\mu}\hat{p}_{\mu}\right]\psi_{m}(x)=0
\end{equation}

\noindent A general state can be an arbitrary superposition of different
mass plane waves. The ``on-mass-shell'' subset of wave functions
\textbf{\uline{do not}} form a complete set for the Hilbert space
with inner product 

\noindent 
\begin{equation}
\left\langle \phi|\psi\right\rangle =\int\phi^{*}(x)\psi(x)d^{4}x
\end{equation}

\noindent For completeness, off-mass-shell states must be included.

\noindent \medskip{}

\noindent \textbf{Bohmian description of the plane wave case}
\begin{equation}
\psi(x)=Ae^{ip^{\mu}x_{\mu}/\hbar}
\end{equation}

\noindent and so the Bohm-Hamilton-Jacobi function is, from \prettyref{eq:Bohm Hamilton-Jacobi function}
(up to an irrelevant additive function of s)

\noindent 
\begin{equation}
S_{B}(x)=p^{\mu}x_{\mu}
\end{equation}

\noindent In this case, the velocity of the Bohmian trajectory is
simply given by

\noindent 
\begin{equation}
u^{\mu}=\frac{dx^{\mu}(s)}{ds}=\frac{\partial^{\mu}S_{B}}{M}=\frac{p^{\mu}}{M}
\end{equation}

\noindent and the Bohmian trajectory is simply given by $x^{\mu}(s)=u^{\mu}s+x^{\mu}(0)$.
If the particle is on the mass shell, then $u^{\mu}u_{\mu}=1,$ otherwise
not. The plane wave is not normalizable, and so it's not a proper
state for the Hilbert space. Note that the trajectory is single valued
here as a function of s. 

\noindent \medskip{}

\noindent \textbf{Gaussian wave functions and double-valued trajectories}

\noindent Holland \cite{holland_quantum_1995} (in section 4.7) derives
the Bohmian trajectory for a Gaussian wave function. We generalize
his calculation to covariant Bohmian mechanics. For the non-relativistic
free particle Schrödinger equation, $i\hbar\frac{\partial\Psi}{\partial t}=-\frac{\hbar^{2}}{2m}\nabla^{2}\Psi,$
starting with the initial (t=0) wave function

\noindent 
\begin{equation}
\Psi_{S}(\overrightarrow{x},0)=\left(2\pi\sigma_{0}^{2}\right)^{-3/4}exp\left(i\overrightarrow{k}\cdot\overrightarrow{x}-\overrightarrow{x}^{2}/\left(4\sigma_{0}^{2}\right)\right)
\end{equation}

\noindent defining $\overrightarrow{u}=\frac{\hbar\overrightarrow{k}}{m}$
one obtains for time t

\noindent 
\begin{equation}
\Psi_{S}(\overrightarrow{x},t)=\left(2\pi s_{t}^{2}\right)^{-3/4}exp\left(i\overrightarrow{k}\cdot\left(\overrightarrow{x}-\frac{1}{2}\overrightarrow{u}t\right)-\left(\overrightarrow{x}-\overrightarrow{u}t\right)^{2}/\left(4s_{t}\sigma_{0}\right)\right)\label{eq:Psi(x,t) for Gaussian}
\end{equation}

\noindent 
\begin{equation}
s_{t}=\sigma_{0}\left(1+i\hbar t/2m\sigma_{0}^{2}\right)
\end{equation}

\noindent The Bohmian action function is then

\noindent %

\noindent 
\begin{equation}
S_{B}(\overrightarrow{x},t)=f(t)+\hbar\overrightarrow{k}\cdot\left(\overrightarrow{x}-\frac{1}{2}\overrightarrow{u}t\right)+\frac{\left(\overrightarrow{x}-\overrightarrow{u}t\right)^{2}\hbar^{2}t/2m\sigma_{0}^{2}}{4\sigma_{0}^{2}\left(1+\hbar^{2}t^{2}/4m^{2}\sigma_{0}^{4}\right)}
\end{equation}

\noindent where $f(t)$ doesn't depend on $x$ and plays no role in
determining the trajectory whose equation is

\noindent 
\begin{equation}
\frac{d\overrightarrow{x}(t)}{dt}=\frac{1}{m}\nabla S_{B}(\overrightarrow{x},t)=\overrightarrow{u}+\frac{\left(\overrightarrow{x}-\overrightarrow{u}t\right)\hbar^{2}t}{4m^{2}\sigma_{0}^{4}\left(1+\hbar^{2}t^{2}/4m^{2}\sigma_{0}^{4}\right)}
\end{equation}

\noindent Holland finds the Bohmian trajectory for this system

\noindent 
\begin{equation}
\overrightarrow{X}_{B}(t)=\overrightarrow{u}t+\overrightarrow{X}_{0}\left(1+\left(\hbar t/2m\sigma_{0}^{2}\right)^{2}\right)^{1/2}\label{eq:Holland trajectory}
\end{equation}

\noindent %

\noindent %

\noindent Next we generalize these results to the covariant Bohmian
case. We have separability in the 4 Minkowski coordinate variables

\noindent 
\begin{equation}
\Psi_{CB}(x,s)=\prod_{\mu=0}^{3}\psi_{\mu}(x^{\mu},s)\label{eq:separable wave function}
\end{equation}

\noindent Despite this notation, $\varPsi_{\mu}$ is not a Lorentz
vector, but rather an indexed set of functions for a separation of
variables. We will have a solution if the $x^{0}$ dependent wave
function in this product satisfies the free particle equation (identical
to the Schrödinger equation)

\noindent 
\begin{equation}
i\hbar\frac{\partial\psi_{0}(x^{0},s)}{\partial s}=-\frac{\hbar^{2}}{2M}\frac{\partial^{2}}{\partial x^{0\,2}}\psi_{0}(x^{0},s)
\end{equation}

\noindent while simultaneously the $x^{j}$ dependent functions satisfy
the complex conjugate equation.

\noindent 
\begin{equation}
i\hbar\frac{\partial\psi_{j}(x^{j},s)}{\partial s}=+\frac{\hbar^{2}}{2M}\frac{\partial^{2}}{\partial x^{j\,2}}\psi_{j}(x^{j},s),\:j\in\left\{ 1,2,3\right\} 
\end{equation}

\noindent Using Holland's results, choose an initial (i.e. at s=0)
wave function of the form \prettyref{eq:separable wave function}
with factors (no sum over $\mu$ here):

\noindent 
\begin{equation}
\psi_{\mu}(x^{\mu},0)=e^{ik^{\mu}x_{\mu}}\left(2\pi\sigma_{I\mu}^{2}\right)^{-1/4}exp\left(-\left(x^{\mu}\right)^{2}/\left(4\sigma_{I\mu}^{2}\right)\right),\:\mu\in\left\{ 0,1,2,3\right\} 
\end{equation}

\noindent There are eight free real parameters $k^{\mu}$ and $\sigma_{I\mu}>0$,
where $u^{\mu}=\frac{\hbar k^{\mu}}{M}$. These wave functions are
normalized in 4D. At other values of $s$ we have by analogy with
the non-relativistic case \prettyref{eq:Psi(x,t) for Gaussian}

\noindent %

\noindent %

\noindent 
\begin{equation}
\begin{array}{cc}
\psi_{0}(x^{0},s)=e^{ik^{0}\left(x_{0}-\frac{1}{2}u_{o}s\right)}\left(2\pi\Sigma_{0}(s)^{2}\right)^{-1/4}\qquad\qquad\qquad\\
\qquad\qquad\times exp\left(-\left(x^{0}-u^{0}s\right)^{2}/\left(4\Sigma_{0}(s)\sigma_{I0}\right)\right)
\end{array}
\end{equation}

\noindent 
\begin{equation}
\begin{array}{cc}
\psi_{j}(x^{j},s)=e^{ik^{j}\left(x_{j}-\frac{1}{2}u_{j}s\right)}\left(2\pi\Sigma_{j}(s)^{2}\right)^{-1/4}\qquad\qquad\qquad\\
\qquad\qquad\times exp\left(-\left(x^{j}-u^{j}s\right)^{2}/\left(4\Sigma_{j}(s)\sigma_{Ij}\right)\right),\:j\in\left\{ 1,2,3\right\} 
\end{array}
\end{equation}

\noindent %

\noindent where

\noindent 
\begin{equation}
\Sigma_{0}(s)=\sigma_{I0}\left(1+i\hbar s/M\sigma_{I0}^{2}\right)
\end{equation}

\noindent 
\begin{equation}
\Sigma_{j}(s)=\sigma_{Ij}\left(1-i\hbar s/M\sigma_{Ij}^{2}\right),\:j\in\left\{ 1,2,3\right\} 
\end{equation}

\noindent it is convenient to also define

\noindent 
\begin{equation}
\sigma_{\mu}(s)=\left|\Sigma_{\mu}(s)\right|=\sigma_{I\mu}\left(1+\left(\hbar s/2M\sigma_{I\mu}^{2}\right)^{2}\right)^{1/2}
\end{equation}

\noindent Although we use subscript notation for notational efficiency,
this does not imply that $\sigma_{I\mu}$, $\sigma_{\mu}$, and $\Sigma_{\mu}(s)$
transform as Lorentz vectors. 

\begin{equation}
\begin{array}{c}
\psi_{CB}(x,s)=e^{ik^{\alpha}\left(x_{\alpha}-u_{\alpha}s/2\right)}\\
\times\prod_{\alpha=0}^{3}\left(2\pi\Sigma_{\alpha}(s)\right)^{-1/4}exp\left(-\left(x^{\alpha}-u^{\alpha}s\right)^{2}/\left(4\Sigma_{\alpha}(s)\sigma_{I\alpha}\right)\right)
\end{array}
\end{equation}

\noindent and the Bohmian action \prettyref{eq:Bohm Hamilton-Jacobi function}
is

\noindent %

\noindent 
\begin{equation}
S_{B}(x,s)=f(s)+\hbar k^{\alpha}\left(x_{\alpha}-u_{\alpha}s/2\right)-\hbar\sum_{\alpha=0}^{3}\left(x^{\alpha}-u^{\alpha}s\right)^{2}\frac{Imag\left(\Sigma_{\alpha}^{*}(s)\right)}{4\sigma_{I\alpha}\Biggl|\Sigma_{\alpha}(s)\Biggr|^{2}}
\end{equation}

\noindent where $f(s)$ is a function only of s. These wave functions
are not eigenstates for mass, energy, or momentum in general. If the
$\sigma_{I\mu}$ values are all large, then the variance in the mass,
energy, and momentum are all small, and so we can approximate an eigenstate
of the mass with this. If in addition $u^{\mu}u_{\mu}=1$ then the
particle is nearly an on-mass-shell eigenstate. Continuing by analogy
with Holland's analysis \cite{holland_quantum_1995}, we obtain the
Bohmian trajectories 

\noindent 
\begin{equation}
X_{B}^{\mu}(s)=u^{\mu}s+X_{B}^{\mu}(0)\left(1+\left(\hbar s/2M\sigma_{I\mu}^{2}\right)^{2}\right)^{1/2}\label{eq:Gaussian Bohm trajectory}
\end{equation}

\noindent and the s-velocity fields are given by

\begin{equation}
\begin{array}{c}
V_{B}^{\,\mu}(s)=u^{\mu}+X_{B}^{\mu}(0)\frac{s\left(\hbar/2M\sigma_{I\mu}^{2}\right)^{2}}{\left(1+\left(\hbar s/2M\sigma_{I\mu}^{2}\right)^{2}\right)^{1/2}}\\
=u^{\mu}+\frac{\left(X_{B}^{\mu}(s)^{\mu}-u^{\mu}s\right)\hbar^{2}s/4M^{2}\sigma_{I\mu}^{4}}{1+\left(\hbar s/2M\sigma_{I\mu}^{2}\right)^{2}}
\end{array}
\end{equation}

\noindent %

\noindent The mass is variable along this trajectory. It is given
by the formula

\noindent 
\begin{equation}
m^{2}(s)=M^{2}V_{B}^{\,\mu}(s)V_{B\,\mu}(s)
\end{equation}

\noindent Notice that the mass depends on $X_{B}^{\mu}(0)$, and that
if this 4-vector is spacelike and very large, then $m^{2}(s)$ can
become negative in principle. However, plugging plausible numbers
into these formulas, this only occurs when $X_{B}^{\mu}(0)$ is many
standard deviations from the mean, and I will ignore this complication
here for the sake of brevity. 

\noindent %

\noindent The trajectory function $X_{B}^{\mu}(s)$ is double-sheeted
as a locally analytic function of a complex variable $s$ due to the
square root factor in \prettyref{eq:Gaussian Bohm trajectory}. We
may write, for the trajectory function on the two Riemann sheets

\noindent 
\begin{equation}
X_{B\pm}^{\mu}(s)=u^{\mu}s\pm X_{B}^{\mu}(0)\left(1+\left(\varGamma(\mu)\right)^{2}s^{2}\right)^{1/2},\:\varGamma(\mu)=\hbar/2M\sigma_{I\mu}^{2}\label{eq:Two trajectories}
\end{equation}

\noindent 
\begin{equation}
V_{B\pm}^{\,\mu}(s)=u^{\mu}\pm X_{B}^{\mu}(0)\frac{s\left(\varGamma(\mu)\right)^{2}}{\left(1+\left(\varGamma(\mu)\right)^{2}s^{2}\right)^{1/2}}\label{eq:Two velocities of Bohm}
\end{equation}

\noindent The asymptotic behavior at large s is

\noindent 
\begin{equation}
V_{B\pm}^{\,\mu}(s)=u^{\mu}\pm X_{B}^{\mu}(0)\varGamma(\mu)+O(X_{B}^{\mu}(0)/\varGamma(\mu)s^{2})\label{eq:asymptotic velocity}
\end{equation}

\noindent 
\begin{equation}
X_{B\pm}^{\mu}(s)=s\left(u^{\mu}\pm\varGamma(\mu)X_{B}^{\mu}(0)\right)+O(X_{B}^{\mu}(0)/\varGamma(\mu)s)
\end{equation}

\noindent the branch points are at $s=\pm i/\varGamma(\mu)$. The
branch cut may be chosen to lie on any arbitrary curve joining these
two branch points. Each of the two trajectories are timelike if $X_{B}^{\mu}(0)$
is close enough to the origin. For extremely large values of $X_{B}^{\mu}(0)$,
the resulting trajectory can also be tachyonic or backward in time,
depending on the orientation of $X_{B}^{\mu}(0)$. To set the scale
for typical atomic problems, consider the value of $\varGamma$ for
an electron with $\sigma_{I\mu}=1\textrm{Å}$ and $X_{B}^{\mu}(0)=1\textrm{Å}$.
We find

\noindent 
\begin{equation}
\varGamma(\mu)=5.7810^{15}s^{-1},\:and\:X_{B}^{\mu}(0)\varGamma(\mu)=0.0019c\label{eq:electron typical values}
\end{equation}
So, the velocities added by the quantum mechanical force are non-relativistic
for this case. These two families of curves are timelike congruences,
although they are not geodesics. In the case where all four $\varGamma(\mu)$
are equal, then they have a time dependent expansion, with the shear
and twist both being zero in the language of timelike congruences
\cite{poisson_relativists_2007}. 

Both Bohmian trajectories in \prettyref{eq:Two trajectories} have
acceleration. But, averaging the momentum of these two together yields

\noindent 
\begin{equation}
P^{\mu}=\frac{1}{2}M\left(V_{B+}^{\,\mu}(s)+V_{B-}^{\,\mu}(s)\right)=Mu^{\mu}
\end{equation}

\noindent which is a conserved value, and independent of $\:X_{B}^{\mu}(0)$,
just as one would expect for a classical free particle. Moreover,
the angular momentum for the pair of trajectories about the center
point of the Gaussian at world time $s=0$ is

\noindent 
\begin{equation}
M_{\pm}^{\mu\nu}(s)=X_{B\pm}^{[\mu}V_{B\pm}^{\nu]}
\end{equation}

\noindent 
\begin{equation}
M^{\mu\nu}(s)=\frac{1}{2}\left[X_{B+}^{[\mu}V_{B+}^{\nu]}+X_{B-}^{[\mu}V_{B-}^{\nu]}\right]=0
\end{equation}

\noindent So the combined orbital angular momentum of the pair of
trajectories equally weighted is zero in this case as well.

\noindent \medskip{}

\noindent \textbf{Why the double-valued solution is more general than
just this one case}

\noindent Consider the Hamilton-Jacobi equation for Bohm action function
$S_{B}$

\noindent 
\begin{equation}
\frac{dS_{B}}{ds}+\frac{1}{2M}\partial_{\mu}S_{B}\partial^{\mu}S_{B}-\frac{\hbar^{2}}{2M}\frac{\partial_{\mu}\partial^{\mu}R}{R}+V=0
\end{equation}

\noindent Solve for one of the derivatives we find a double valued
solution

\begin{equation}
\begin{array}{c}
\partial^{\alpha}S_{B}=\pm\Biggl(-\eta^{\alpha\alpha}\sum_{\mu\neq\alpha}\partial_{\mu}S_{B}\partial^{\mu}S_{B}\\
+\eta^{\alpha\alpha}2M\left(\frac{dS_{B}}{ds}+\frac{1}{2M}\partial_{\mu}S_{B}\partial^{\mu}S_{B}-\frac{\hbar^{2}}{2M}\frac{\partial_{\mu}\partial^{\mu}R}{R}+V\right)\Biggr)^{1/2}
\end{array}
\end{equation}

\noindent %

\noindent In general, the two Riemann sheets evidenced by the $\pm$
sign can be visited by analytic continuation in the complex planes
of the arguments of $S_{B}$. There is at least one exception. If
the wave function is a perfect plane wave, then $\partial_{j}S_{B}$
is a constant, and has only one value. So, in general we can expect
multi-valued trajectories in Bohmian mechanics in complex space time.
In the case of free particles it is plausible then to expect the double-sheeted
solutions of the form $\partial_{\mu}S_{B}(x,s)=Mu_{\mu}\pm F_{\mu}(x,s)$
for some vector function $F_{\mu}$. There may be exceptions. The
plane wave is an exception, but since it's not normalizable it's excluded
from consideration.

\noindent \medskip{}

\section{Generalized analytic continuation by partition of unity}

\noindent The possibility that multiple null roots contribute to solutions
of the Liénard-Wiechert potentials suggests that when this occurs
the solutions to Maxwell's equation are not unique. This was pointed
out in \cite{davidson_lorentz-dirac_2012} where a new form of generalized
analytic continuation (GAN) was introduced. In that study, the trajectory
function was an entire function of the complex proper time, although
the retarded time was multivalued. In the present paper, we have found
that the Bohmian trajectory function itself is not single valued as
a function of complex world time $s$. This provides us with the option
to apply GAN to the trajectories before calculating the set of null
times. First we define exactly what is meant by GAN. Let $f(z,W)$
be a locally analytic function of a single complex variable z with
a discrete set of branch points, poles, and essential singularities,
and let $W$ be a set of complex parameters that $f$ depends on in
a smooth and locally analytic way as well, and contained in some complex
manifold. From the values of this function in any neighborhood of
a point in the complex z plane, the function can be derived everywhere
else in z by analytic continuation. If the function is multivalued,
then this will be discovered by examining all analytic continuation
paths. In general the function will have multiple Riemann sheets.
Let their number be $N$, although $N$ could be infinite. N could
depend on $W$, but we can take it to be an upper bound in this case
without loss of generality. Now consider the following construction
of a complex partition of unity. Let the $P_{i}(W)$ be smooth and
analytic complex functions of the parameters W which satisfy the condition

\noindent 
\begin{equation}
\sum_{i=1}^{N}P_{i}(W)=1,\:
\end{equation}
Where $N$ is the number of Riemann sheets of $f(z)$. Let branch
cuts be selected which define the Riemann sheets uniquely. Next, consider
a locally analytic point z which does not lie on any branch cut. Then,
there exists a neighborhood $\mathbf{\mathcal{\mathscr{\mathcal{N}}}}$
about z which does not intersect any branch cut. We define a new function.
We start by doing nothing in the following form

\noindent 
\begin{equation}
f(z)=\sum_{i=1}^{N}P_{i}(W)f(z)
\end{equation}
and next we create an intermediary function of $N$ variables $z_{i}$ 

\noindent 
\begin{equation}
F(z_{1},...,z_{N})=\sum_{i=1}^{N}P_{i}(W)f(z_{i}),\:z_{i}\in\mathcal{\mathscr{\mathcal{N}}}
\end{equation}
and finally we analytically continue each $z_{i}$ along a different
path to arrive back at the starting value of $z$, but ending up on
the ith Riemann sheet for the function $f$. In doing this, the paths
for the various $z_{i}$ will leave the local neighborhood $\mathbf{\mathcal{\mathscr{\mathcal{N}}}}$
temporarily as they must pass through branch cuts in order to reach
other Riemann sheets. But eventually all the paths return to the starting
point $z$.

\noindent 
\begin{equation}
F(z_{1},...,z_{N})\rightarrow\sum_{i=1}^{N}P_{i}(W)f_{i}(z_{i})\rightarrow\sum_{i=1}^{N}P_{i}(W)f_{i}(z)=G(z,\left\{ P_{i}(W)\right\} )
\end{equation}
This is what we mean by generalized analytic continuation or GAN.
The new function $G(z,\left\{ P_{i}(W)\right\} )$ depends on the
arbitrary weighting parameters $P_{i}(W)$. So there is a great proliferation
of one function into a whole family of GANs. Let us name this mapping
$GAN:f(z)\rightarrow a\:subset\:of\:all\:locally\:analytic\:functions$. 

\noindent \textbf{A note of caution }The GAN continuation of a sum
is not necessarily the same as the sum of the GANs. Some restrictions
are required to avoid contradictions. For example, the function f(z)=0
is single-valued, and we must insist that $GAN$ on this function
yields just the zero function back again. But we can write $0=\sqrt{z}-\sqrt{z}$,
and therefore if we assume linearity by writing $GAN(0)=GAN(\sqrt{z})+GAN(-\sqrt{z})$,
where the addition would be the set of pairwise sums of all pairs
of functions $f_{1}+f_{2}$ where $f_{1}\in GAN(f_{1})$ and $f_{2}\in GAN(f_{2})$.
Obviously $GAN(\sqrt{z})+GAN(-\sqrt{z})$ contains many more functions
than the zero function. So in general we cannot assume linearity for
GAN. But there are some cases where linearity does not cause a problem.
If one or both of the functions $A(z)$ and $B(z)$ are single valued,
then we can safely write $GAN(A+B)=GAN(A)+GAN(B)$. But in general
we cannot. Similarly for multiplication. If $f(z)=A(z)B(z)$, then
in general $GAN(f)=GAN(AB)\neq GAN(A)GAN(B)$ unless one (or both)
of $\left\{ A,B\right\} $ is an entire function in which case equality
does hold. By $GAN(A)GAN(B)$ is meant just pairwise multiplication
of the elements of $GAN(A)$ with those of $GAN(B)$.

\noindent \textbf{Examples}

\noindent For entire analytic functions, like $exp(z)$ or any polynomial
in $z$, the GAN continuation of the function gives just that single
function back. There must be more than one Riemann sheet to give non-trivial
results.

\noindent \textbf{Lemma} $GAN(\sqrt{z})$ is the set of functions
$X\sqrt{z}$, where $X$ is an arbitrary complex number.

\noindent \textbf{Proof} $\sqrt{z}$ has two Riemann sheets. Let $P_{1}+P_{2}=1$
with $P_{1}$ and $P_{2}$ complex numbers. The branch point is at
$z=0$. We can draw the branch cut from this point to $\infty$ any
way we choose. Once the branch cut is defined, the two Riemann sheets
are then defined. It doesn't matter how we draw the branch cut and
define the two Riemann sheets. The function $\sqrt{z}$ on the first
Riemann sheet is the negative of its value on the second sheet. $f_{1}(z)=-f_{2}(z)$.
Let z be a point on Riemann sheet 1. $GAN(\sqrt{z})$ is therefore
$P_{1}\sqrt{z}-P_{2}\sqrt{z}=\left(P_{1}-P_{2}\right)\sqrt{z}$, but
$P_{1}-P_{s}$ can take on any complex value. This function can be
extended analytically in z, and we get by so doing the original double-valued
$\sqrt{z}$ function, but multiplied by an arbitrary factor $\left(P_{1}-P_{2}\right)$,
and so the lemma is proven.

\noindent In like manner, it can be shown that the GAN of $z^{1/n}$
is $Xz^{1/n}$, and of $ln(z)$ is $X+ln(z)$ for any complex $X.$

\noindent Now consider the family of Bohmian trajectories in our free
particle example \prettyref{eq:Gaussian Bohm trajectory}. 

\noindent 
\begin{equation}
X_{B}^{\mu}(s)=u^{\mu}s+X_{B}^{\mu}(0)\left(1+\left(\hbar s/2M\sigma_{I\mu}^{2}\right)^{2}\right)^{1/2}
\end{equation}

\noindent We require that $X_{B}^{\mu}(0)\neq0$ for $\mu=0,..,3.$
Note that $u^{\mu}s$ is single valued. In this case

\noindent 
\begin{equation}
GAN(X_{B}^{\mu}(s))=u^{\mu}s+X_{B}^{\mu}(0)GAN\left(\left(1+\left(\hbar s/2M\sigma_{I\mu}^{2}\right)^{2}\right)^{1/2}\right)
\end{equation}

\noindent This gives the set of all trajectories of the form

\noindent 
\begin{equation}
GAN(X_{B}^{\mu}(s))=\left\{ u^{\mu}s+A^{\mu}\left(1+\left(\hbar s/2M\sigma_{I\mu}^{2}\right)^{2}\right)^{1/2}\forall A^{\mu}\in\mathbb{C}\right\} \label{eq:GAN XB 1}
\end{equation}

\noindent The Bohm trajectories, are the subset of these where $A^{\mu}$
are real. 

\noindent 
\begin{equation}
Re(GAN(X_{B}^{\mu}(s)))=\left\{ Bohmian\:trajectories\right\} \label{eq:GAN XB 2}
\end{equation}

\noindent Where $Re(GAN(X_{B}^{\mu}(s)))$ is the subset of real trajectories.
So we see that a single trajectory determines the whole set of Bohmian
trajectories by the generalized analytic continuation procedure. 

\textcolor{black}{We see that all the Bohmian trajectories in this
case belong to a GAN starting from a single seed trajectory which
satisfies $X_{B}^{\mu}(0)\neq0$ for any value of $\mu=\left\{ 0,1,2,3\right\} $.
In other words if you take one of the Bohmian trajectories, you can
derive all the other ones from it using the GAN procedure without
knowing in advance what the wave function was. This result has been
shown for the special case of the Gaussian wave function only. Whether
it is a general result of the Bohmian theory for free particles is
not known. Once you have analytic expressions for all the trajectories
as functions of world time, you can then differentiate them to find
the velocity and acceleration of these classical trajectories. If
the external potential force is known, one can then calculate the
quantum mechanical potential force for all of the trajectories as
a function of position and world time. Using the velocities, one can
also calculate Hamilton's principle function. Knowing both the quantum
mechanical force and Hamilton' principle function, one should be able
to figure out a class of equivalent wave functions that will give
this Hamilton-Jacobi equation using Bohmian mechanics rules. Thus
a single trajectory, in the cases studied in this paper at least,
has hidden in its analytic structure information about all the other
trajectories, and one never has to introduce wave mechanics as a separate
postulate. This follows from the GAN procedure as outlined above.
The single starting trajectory for this procedure is a purely particle
picture, but the result is that a wave equation can be deduced from
just this one trajectory which then generates all the other trajectories
by the Bohmian mechanics rules. Proving this in general may be difficult,
and it may not be generally true. I don't know how extensively these
results can be extended to other cases. In the next few sections the
electromagnetic fields in complex space-time are examined, and we
shall see how this whole family of trajectories can contribute to
the electromagnetic field produced by a charged particle moving along
a single trajectory in this way. As far as the electromagnetic field
produced is concerned, the source acts as a set or ensemble of trajectories,
not just a single one. These conclusions are coming from a classical
field theory, except that we put in the Bohmian trajectory by fiat.
Everything being described here is therefore a classical phenomenon,
yet it's very similar to aspects of quantum mechanics. The key ingredients
are complex space-time, the Riemann sheet structure of multivalued
functions, and the generalized analytic continuation.}

\noindent %

\section{\label{sec:Electromagnetism-in-complex}Electromagnetism in complex
space-time}

\noindent The Kerr-Newman particle is modeled as a point charge which
is slightly displaced from the real hypersurface in complex Minkowski
space $(\mathbb{C}M^{\text{4}})$\cite{newman_metric_1965,newman_maxwells_1973,newman_complex_1973,newman_heaven_1976,burinskii_kerr_2007,burinskii_dirac_2008,lynden-bell_magic_2003,pekeris_electromagnetic_1987}.
The complex theory makes contact with reality by ultimately considering
the electromagnetic fields ``projected'' onto the real space-time
hyperspace. There are two equivalent formalisms that one can use.
The Riemann-Silberstein complex vector field can be used \cite{lynden-bell_magic_2003}

\noindent 
\begin{equation}
\mathbf{\mathrm{\overrightarrow{W}}}=\mathbf{\mathrm{\overrightarrow{E}}}+i\overrightarrow{\mathbf{\mathrm{B}}}\label{eq:Riemann-Silberstein}
\end{equation}
or the equivalent covariant form \cite{newman_maxwells_1973}

\noindent 
\begin{equation}
W^{\mu\nu}=F^{\mu\nu}+i\,^{*}\!F^{\mu\nu},\:^{\ast}\!W^{\mu\nu}=-iW^{\mu\nu}\label{eq:Complex Faraday}
\end{equation}
where $^{\ast}\!W$ is the dual of $W$. From \prettyref{eq:Complex Faraday}
we see that $W^{\mu\nu}$ is anti self-dual (ASD). The electromagnetic
energy density and Poynting vector are ($\overline{{\normalcolor \mathbf{\mathbf{\mathrm{\overrightarrow{W}}}}}}=CC\:of\:\mathbf{\mathbf{\mathrm{\overrightarrow{W}}}}=\mathbf{\mathrm{\overrightarrow{W}}}^{*})$

\noindent 
\begin{equation}
\mathcal{E}_{el}=\frac{1}{2}\overline{{\normalcolor \mathbf{\mathbf{\mathrm{\overrightarrow{W}}}}}}\cdot\mathbf{\mathbf{\mathrm{\overrightarrow{W}}}},\:\mathbf{P}_{el}=\frac{i}{2}\mathbf{\mathbf{\mathrm{\overrightarrow{W}}}}\times\overline{\mathbf{\mathbf{\mathrm{\overrightarrow{W}}}}}\label{eq: energy and momentum for RS vector}
\end{equation}
The stress energy tensor on the real hyperspace is a function of the
real-valued physical Faraday tensor 

\noindent 
\begin{equation}
4\pi T^{\mu\nu}=F_{phys\:\lambda}^{\mu}F_{phys}^{\lambda\nu}+\frac{1}{4}g^{\mu\nu}F_{phys\alpha\beta}F_{phys}^{\alpha\beta}\label{eq: Stress energy tensor}
\end{equation}
\textcolor{black}{and this can be expressed in terms of the complex
Faraday tensor with the substitution }

\noindent 
\begin{equation}
F_{phys}^{\mu\nu}(x)=Re\:W^{\mu\nu}(x),\:for\:all\:x^{\alpha}\:real\label{eq: Physical Faraday tensor}
\end{equation}
This simple prescription gets us back to real fields on real space-time.
The absence of magnetic charge requires that $F_{phys}$ satisfies
the electromagnetic Bianchi identity on real space-time, i.e. $\partial_{\mu}\left(^{\ast}\!F_{phys}\right)^{\mu\nu}=0$. 

The static Kerr-Newman particle is modeled as a point charge located
at a point in complex 3-space at $\mathbf{z_{0}}=\mathbf{x_{0}}+i\mathbf{b}$,
where $\mathbf{x_{0}}$ and $\mathbf{b}$ are real 3-vectors. One
introduces a complex \textcolor{black}{Coloumb} potential created
by this particle

\noindent 
\begin{equation}
\Phi(z)=q/\sqrt{\left(\mathbf{z}-\mathbf{z}_{0}\right)^{2}}\label{eq: Complex_KN_potential}
\end{equation}

\noindent and from this the fields can be derived, and they allow
a metric of the Kerr-Schild type.

\section{Liénard\textendash Wiechert potentials in complex spacetime }

\noindent The Li\'{e}nard-Wiechert potentials are \cite{jackson_classical_1999}

\noindent 
\begin{equation}
A^{\mu}(x)=\left.\frac{qv^{\mu}(s)}{v(s)\cdot(x-z(s))}\,\right|_{s=s_{r}}\label{eq:Lienard-Wiechert potential}
\end{equation}

\noindent Normally only one retarded time contributes to the field
from a single real timelike trajectory. The Faraday tensor from a
single root may be calculated from the formula

\begin{equation}
\begin{aligned}F^{\mu\nu}(x) & =\frac{q}{V_{B}(s)\cdot\left(x-X_{B}(s)\right)}\\
 & \qquad\times\frac{d}{ds}\left[\frac{\left(x-X_{B}(s)\right)^{\mu}V_{B}^{\nu}(s)-\left(x-X_{B}(s)\right)^{\nu}V_{B}^{\mu}(s)}{V_{B}(s)\cdot\left(x-X_{B}(s)\right)}\right]\biggr|_{s=s_{r}}
\end{aligned}
\end{equation}

\noindent Let $x^{\mu}$ be an arbitrary field point in real Minkowski
space where we wish to calculate the electromagnetic field generated
by the moving charged particle. Then the null condition for calculating
the null world times is

\noindent 
\begin{equation}
\left(x^{\mu}-X_{B}^{\mu}(s)\right)\left(x_{\mu}-X_{B\mu}(s)\right)=0
\end{equation}

\noindent Since we are working in complex space-time, we include all
roots for the variable $s$ in the complex plane. This is a major
difference between the complex space-time embedding and the usual
real Minkowski space electromagnetism. The solutions to this equation
are functions of the field point x which we take to be real. The multiple
solutions to this equation will be denoted by $\left\{ s_{n}(x)\right\} $.
If $X_{B}(s)$ is a polynomial of degree N in s, then there will be
$2N$ roots in the complex s plane obtained by standard analytical
continuation. An example of this kind of multiple root situation was
studied in \cite{davidson_lorentz-dirac_2012} where there were an
infinite number of roots. 

If $X_{B\mu}(s)$ is multivalued as a function of $s$, as is the
case for our Bohmian Gaussian example, then we can consider any root
from any of the GAN continuation trajectories as possibly contributing
to the field. We get the most interesting results if we apply GAN
two times, first to the trajectories themselves, and next to the Liénard-Wiechert
potentials. We look for all roots in the complex s plane, and there
can be many such roots. In our case, we have just one retarded root
per trajectory, assuming that only timelike trajectories are included
in the Bohmian trajectory set, and therefore we include only these
roots\footnote{The non-timelike trajectories in the Bohmian set constitute a negligibly
small fraction and are ignored here}. So we now plug in the covariant Bohmian trajectories calculated
above into this formula. For simplicity, let us choose all of the
$\sigma_{I\mu}$ equal to each other. Let $\sigma_{I}$ denote the
common value. Define a value of $\varGamma$ which is independent
of $\mu$ as

\noindent 
\begin{equation}
\varGamma=\frac{\hbar}{2M\sigma_{I}^{2}}
\end{equation}

\noindent The fundamental trajectory equation is then

\noindent 
\begin{equation}
X_{B}^{\mu}(s)=u^{\mu}s+X_{B}^{\mu}(0)\left(1+\varGamma^{2}s^{2}\right)^{1/2}
\end{equation}

\noindent where $X_{B}^{\mu}(0)$ is restricted to real values, and
represents any one of the ensemble of Bohmian trajectories at $s=0$.
So, the root equation for a single trajectory becomes (notation $\left(x\right)^{2}=x^{\mu}x_{\mu}$).

\noindent 
\begin{equation}
\left(x-us-X_{B}(0)\left(1+\varGamma^{2}s^{2}\right)^{1/2}\right)^{2}=0\label{eq:Root equation}
\end{equation}

\noindent Applying a GAN transformation to the trajectory as in \ref{eq:GAN XB 1}
and \ref{eq:GAN XB 2} we obtain 

\begin{equation}
Re\:GAN(X_{B}^{\mu}(s))=\{All\:real\:Bohmian\:trajectories\}
\end{equation}

\noindent Although the GAN transformation produces complex trajectories
as well as real ones, we ignore these by assuming that the weighting
funtion has support only over real trajectories. Therefore we have
one retarded time for each Bohmian trajectory, and they all contribute
in a weighted sum to the electromagnetic field. Denote the solutions
by $\left\{ s_{r}(x,X_{B}(0))\right\} $, and here $X_{B}(0)$ can
be any real 4-vector. Each root solution produces a Liénard-Wiechert
solution, and so we can write

\begin{equation}
F^{uv}(x)=\sum_{X_{B}(0)}P(X_{B}(0))F_{X_{B}(0)}^{\mu\nu}(x,s_{r}(x,X_{B}(0)))\label{eq:GAN transformation on fields}
\end{equation}

\noindent where $F_{X_{B}(0)}^{\mu\nu}(x,s_{r})$ is the field at
point $x$ produced by a particle of charge $Q$ moving along a trajectory
that passes through $X_{B}(0)$ , and $P(X_{B}(0))$ is the weight
given to the trajectory that passes through $X_{B}(0)$ normalized
so that\footnote{Q here is the actual physical charge of the particle}

\begin{equation}
\sum_{X_{B}(0)}P(X_{B}(0))=1
\end{equation}

\noindent The summation in \prettyref{eq:GAN transformation on fields}
is just a GAN transformation on the multivalued Faraday tensor. The
solutions to Maxwell's equations in this case are simply not unique.
In general the $P(X_{B}(0))$ can be arbitrary complex numbers, but
the natural choice for these is the Bohmian weighting which generates
Born's rule, which becomes an integration over all starting positions
$X_{B}(0)$ restricted to real values, and weighted by their probability
density. $\rho=\Psi^{*}\Psi$ is the probability density. We can express
it in terms of Bohmian trajectories

\begin{equation}
\begin{array}{c}
\rho(x,s)=\sum_{X_{B}(0)}P(X_{B}(0))\delta^{4}(x-X_{B}(s))\\
=\int d^{4}X_{B}(0)\rho(X_{B}(0),0)\delta^{4}(x-X_{B}(s))=\Psi^{*}(x,s)\Psi(x,s)
\end{array}
\end{equation}

\noindent and the integrated probability current is $J^{\mu}(x)=\int_{-\infty}^{\infty}V^{\mu}(x,s)\varrho(x,s)ds$
as in \prettyref{eq:Bohmian current density}. The field is an ensemble
average over all these trajectories

\begin{equation}
F^{uv}(x)=\int d^{4}X_{0}\varrho(X_{0})F_{X_{0}}^{\mu\nu}(x,s_{r}(x,X_{0}))
\end{equation}

\noindent The result is that the collective current density of the
ensemble is proportional to the probability current calculated from
the quantum wave equation. This is then equivalent to 

\begin{equation}
\partial_{\mu}F^{uv}(x)=\mathcal{J_{\mathcal{EM}}^{\nu}}(x)=\frac{\hbar q}{2Mi}\int_{-\infty}^{\infty}ds\varPsi^{*}(x,s)\overleftrightarrow{\partial^{\nu}}\varPsi(x,s)\label{eq:extended quantum current}
\end{equation}

\noindent and we see that the classical electromagnetic field can
be produced by the coherent sum of all the Bohmian trajectories, and
so it can look as if it's produced by an extended object. The reader
should note that the integral of $\mathcal{J_{\mathcal{EM}}^{\mathrm{0}}}(x)$
over 3-space is independant of $x^{0}$ because of \prettyref{eq:Particle continuity equation}
and the value of the parameter $q$ must be chosen so that this value
gives the total charge of the particle. In words, a single charged
particle trajectory generates a field whose generating current is
that of an extended object. This is due to the non-unique solutions
to the Liénard-Wiechert potentials when embedded in complex space-time,
along with the particular solutions to the Bohmian trajectories which
provide the mathematical justification, since any single one (excluding
a set of measure zero) can be analytically continued (through GAN)
to any other one, and thus a continuum of retarded times and positions
are coherently added together yielding the resultant electromagnetic
field of an extended object. This is an interesting result, but it
should be remembered that it depends on applying GAN to the trajectories
first to obtain an infinite set of them, and then applying it again
to the Faraday tensor to produce a weighted sum over the multitude
of retarded times so obtained. It is not obvious that solutions obtained
in this way should be admitted in an electromagnetic theory in complex
space-time. But if they are allowed, then we see that we can potentially
obtain a solution to the ancient problem of the infinite electrostatic
self-energy of classical point particles.

\section{The radiation problem in the full-relativistic case for a free particle}

\noindent The GAN continuation result provides a nice explanation
for understanding why the free-particle Bohm trajectory need not radiate.
Using results from \cite{davidson_quantum_2007} we see first of all
that the Klein-Gordon current of a positive energy free-particle wave
function are non-radiating electromagnetically. To make use of this,
we can write any free-particle wave function $\psi$ as a sum over
fixed mass eigenstates 

\noindent 
\begin{equation}
\varPsi(x,s)=\int_{-\infty}^{\infty}dm\psi_{m}(x)e^{-ism^{2}/2M\hbar},\:\left[m^{2}-\hat{p}^{\mu}\hat{p}_{\mu}\right]\psi_{m}(x)=0
\end{equation}

\noindent Now using \prettyref{eq:extended quantum current} , we
can write the electromagnetic 4-current as

\noindent 
\begin{equation}
\mathcal{J_{\mathcal{EM}}}(x)=\frac{\pi\hbar^{2}q}{i}\int_{-\infty}^{\infty}dm\frac{1}{m}\varPsi_{m}^{*}(x)\overleftrightarrow{\partial^{\mu}}\varPsi_{m}(x)\label{eq:real quantum current}
\end{equation}

\noindent where $q$ is the charge up to a normalization factor, as
explained in \ref{eq:Charge Normalization I} and \ref{eq:Charge normalization II}.
It is necessary to assume that $\Psi_{m}=0$ for $m<\epsilon$ for
some positive $\epsilon$. I assume that the $\varPsi_{m}(x)$ are
built up from positive energy plane waves which is a very good approximation
to our Gaussian wave functions. Now, if two non-radiating currents
are added together, then the sum is also a non-radiating current \cite{davidson_quantum_2007}.
Extending this argument, we see that $\mathcal{J_{\mathcal{EM}}}(x)$
can be a non-radiating current provided that $\varPsi_{m}(x)$ contains
only positive mass and positive energy plane waves, and that certain
other convergence conditions are satisfied as described in \cite{davidson_quantum_2007}.
Since $\Psi$ falls off as a Gaussian in all four coordinates, I expect
that the convergence conditions will be satisfied. Strictly speaking,
our Gaussian example does have an extremely small contribution from
negative mass states, and so I'm assuming that these can be safely
projected out without affecting the convergence requirements for the
zero radiation proof. 

So, if nature chooses to weight the trajectories by Bohm's prescription,
then we get plausibly at least a non-radiating solution, and by this
argument, we can think of each trajectory as being non-radiating.
The only new ingredient is the idea of a generalized analytic continuation
applied to complex space-time embedding. For the general free-particle
case, it remains an open question. 

\section{The near field for an isolated particle}

\noindent What does the electromagnetic field look like if we are
near to the particle? This question introduces a measurement problem.
How do we measure the near field of a quantum particle? One way is
to scatter another charged particle off of it, and look at deep elastic
scattering. But this involves solving a different quantum wave equation
for the two-body problem. The near field of an isolated particle is
not measurable, because as soon as we try to measure it we must introduce
at least one new charged particle into interaction with it, and therefore
it is no longer isolated. In the far field, we found that for an isolated
charged particle in a particular quantum state the classical electromagnetic
charge density and current could be taken proportional to the probability
4-current derived from the wave function, and that this avoided radiation.
This charge density is spread out over a large volume compared to
say the Compton wavelength of the particle. It's not likely to produce
any singular field points, although there may be exceptions. The point
particle in complex space-time and moving along a Bohmian trajectory
can therefore act like a kind of jellium of extended charge density
in complex space-time. There is no obvious conflict with experiment
if we simply assume that this extended charge model really does describe
the reality of a completely isolated charged particle. Now if we make
this assumption, which is based on the GAN continuation together with
the assignment of the Bohmian weighting factors for each trajectory,
then lo and behold, the classical electromagnetic energy and stress
tensor is that of an extended charge which is no longer definitely
singular, but calculable and probably for most wave functions finite
due to the extended charge density. In other words, this theory offers
a possible resolution to the long-standing problem of how to avoid
the infinite self energy of an isolated point charge. In fact, since
the extended charge density is non-radiating, it is concievable that
the self-force of this charge density will be zero too, and therefore
there may not be any runaway solutions that have plagued classical
electromagnetism since the work of Abraham and Lorentz. I do not dismiss
the Lorentz-Dirac equation though, because it may be the origin of
zitterbewegung as discussed in \cite{davidson_lorentz-dirac_2012}.
So, at least there is a possibility in this theory to resolve the
ancient paradoxes of point particle electrodynamics which have plagued
this theory for some time. These results bring to mind the theory
of self-field electrodynamics by Barut and Dowling \cite{barut_quantum-electrodynamics_1988,barut_self-field_1990,barut_interpretation_1991,barut_qed_2014}.

An experiment has been proposed that could test this jellium picture
of an isolated quantum charged particle. It assumes that in the presence
of a weak classical external electromagnetic field, the extended charge
distribution \prettyref{eq:extended quantum current} is still valid.
This hydrodynamic model leads to possibly testable bremsstrahlung
effects \cite{davidson_predictions_2004}.

\section{In the non-relativistic limit, the radiation fields of the two Riemann
sheets cancel if we average them together}

\noindent In the non-relativistic limit, we can show the lack of radiation
in another way that does not involve the assumption of GAN. In this
limit we have

\noindent 
\begin{equation}
u=\left\{ u^{0},\overrightarrow{u}\right\} ,\:\overrightarrow{u}^{2}\ll\left(u^{0}\right)^{2},\:and\:s_{r+}\approx s_{r-}
\end{equation}

\noindent and therefore from \prettyref{eq:Two trajectories}, \prettyref{eq:Two velocities of Bohm},
and $V_{B-}^{\nu}(s)\approx V_{B+}^{\nu}(s)$ it follows that the
radiation field from just these two sources vanishes if they are weighted
equally.

\noindent 
\begin{equation}
F^{\mu\nu}(x)=\frac{F_{+}^{\mu\nu}(x)+F_{-}^{\mu\nu}(x)}{2}=0+O\left(\frac{1}{R^{2}}\right)
\end{equation}

\noindent And so the total electromagnetic radiation is zero in this
case also, but only standard analytic continuation is required, not
the more radical GAN technique. Complex space-time is still required.
The radiation fields of the two particles cancel one another in the
non-relativistic limit. So, we have a second possible way of explaining
why free Bohmian particles might not radiate, or radiate very much,
in the non-relativistic limit. Both Bohmian trajectories in \prettyref{eq:Two trajectories}
experience non-zero acceleration. However, averaging the momentum
of these two together yields

\noindent 
\begin{equation}
P^{\mu}=\frac{1}{2}M\left(V_{B+}^{\,\mu}(s)+V_{B-}^{\,\mu}(s)\right)=Mu^{\mu}
\end{equation}

\noindent So the average momentum of the two trajectories is simply
a constant, just as one would expect for a classical free particle.
Moreover, the angular momentum for the pair of trajectories about
the center point of the Gaussian at world time $s=0$ is

\noindent 
\begin{equation}
M_{\pm}^{\mu\nu}(s)=X_{B\pm}^{[\mu}V_{B\pm}^{\nu]}=u^{[\mu}X_{B\pm}^{\nu]}(0)\left[\frac{s^{2}\varGamma(\nu)^{2}}{\left(1+\varGamma(\nu)^{2}s^{2}\right)^{1/2}}-\left(1+\varGamma(\mu)^{2}s^{2}\right)^{1/2}\right]
\end{equation}

\noindent 
\begin{equation}
M^{\mu\nu}(s)=\frac{1}{2}\left[X_{B+}^{[\mu}V_{B+}^{\nu]}+X_{B-}^{[\mu}V_{B-}^{\nu]}\right]=0
\end{equation}

\noindent So the combined orbital angular momentum of the pair of
trajectories equally weighted is zero.

\noindent \medskip{}

\noindent \textbf{Riemann sheet structure of N particle system}

\noindent Imagine an N particle system of non-interacting particles
described by \prettyref{eq:Fanchi equation}. Assume for simplicity
that the wave function is a simple product for starters. Then, there
will be in general two Riemann sheet trajectories per particle. This
means that for N particles there will be $2^{N}$ Riemann sheets.
For entangled states the situation is not clear, and will be left
as a future research problem. 

\noindent %

\section{Spin - Modeling a Kerr-Newman particle off the real axis and riding
along a Bohmian trajectory}

\noindent Up to this point we have been considering real trajectories,
and therefore spinless Kerr-Newman particles, or Reissner\textendash Nordström
metric solutions.\footnote{For elementary particles, these metrics have a naked singularity due
to the electromagnetic field energy. If the particle becomes effectively
an extended jellium by the GAN transformation, then this naked singularity
would generally disappear. } The Kerr-Newman particle in complex spacetime, and in a Lorentz frame
in which it is at rest, is displaced from the real axis by a fixed
amount $\overrightarrow{O}=i\overrightarrow{J}/mc$, where $J=\left|\overrightarrow{J}\right|$
is the angular momentum of the Kerr-Newman solution \cite{burinskii_kerr-schild_2000}.
This is the more interesting case, since most elementary charged particles
(quarks and leptons) have spin. In our Gaussian case, if $u^{\mu}u_{\mu}=1$,
then we can move to its rest frame with a single Lorentz boost. So
we can obtain the imaginary part of a moving particle by Lorentz boosting
$O^{\mu}=\left\{ 0,i\overrightarrow{J}/mc\right\} $ to a proper velocity
$u$. $O^{\mu}(u)=\Lambda_{\nu}^{\mu}(u)O^{\nu}$. The natural thing
to do is to add this boosted $O^{\mu}$ to $X_{B\pm}^{\mu}(0)$ in
\prettyref{eq:Two trajectories}, but when we do this we run into
poblems because then the displacement along the imaginary direction
will grow and change with s in a way that will change its spin value.
To see this, recall that for our Gaussian case this would imply

\noindent 
\begin{equation}
X_{B\pm}^{\mu}(s)=u^{\mu}s\pm\left(X_{B}^{\mu}(0)+O^{^{\mu}}(u)\right)\left(1+\varGamma(\mu)^{2}s^{2}\right)^{1/2}\label{eq:Kerr-Newman Bohm trajectory}
\end{equation}

\noindent for large values of s the imaginary term proportional to
$O^{^{\mu}}(u)$ grows linearly in s. The spin angular momentum is
therefore not constant, and also the particle does not remain close
to the real hyperspace. This is inconsistent with the quantization
of spin angular momentum. If $\varGamma(\mu)$ is small enough, and
the particle lifetime is short then the present model might still
be possible, but for an electron, as shown above \prettyref{eq:electron typical values},
the spin would vary with $s$ in a physically unacceptable manner.
Therefore this way of introducing spin cannot describe quantum electrons.
Muons probably can't be described either this way. But quarks have
a chance because they are never free, except in the quark gluon plasma.
Vector Bosons $W^{+},\:W^{-}$ have very short lifetimes, and their
measured g factor is also reasonably close to 2.0 \cite{holstein_how_2006}
as predicted by the Kerr-Newman theory. They are too short-lived for
the variation of the imaginary part of their Bohmian trajectory to
cause a problem.

There is another way to plausibly introduce spin. We can simply assume
that the Bohmian trajectories for a free particle give the real part
of the trajectory only, and that the effect of spin is simply to add
a constant complex term to this real trajectory. This leads to the
same current density as for the spinless case \prettyref{eq:extended quantum current},
with a modified electromagnetic potential. In the usual real case,
the retarded potentials can be written 

\begin{equation}
A^{\mu}(x)=\int\frac{J^{\mu}(\overrightarrow{x}^{\prime},x^{\prime}{}_{r}^{0})}{\sqrt{\left(\overrightarrow{x}-\overrightarrow{x}^{\prime}\right)^{2}}}d^{3}x^{\prime},\;x^{\prime}{}_{r}^{0}=x^{0}-\sqrt{\left(\overrightarrow{x}-\overrightarrow{x}^{\prime}\right)^{2}}
\end{equation}

\noindent Now we would like to analytically continue this by $x^{\prime}\rightarrow x^{\prime}+O$.
Let's start off with a real displacement, so all $O^{\mu}$ are real
initially. Then the vector potential becomes

\begin{equation}
A^{\mu}(x,O)=\int\frac{J^{\mu}(\overrightarrow{x}^{\prime},x^{\prime}{}_{r}^{0})}{\sqrt{\left(\overrightarrow{x}-\overrightarrow{x}^{\prime}-\overrightarrow{O}\right)^{2}}}d^{3}x^{\prime},\;x^{\prime}\,{}_{r}^{0}=x^{0}-O^{0}-\sqrt{\left(\overrightarrow{x}-\overrightarrow{x}^{\prime}-\overrightarrow{O}\right)^{2}}
\end{equation}

\noindent %

\noindent where $x^{\prime}{}_{r}^{0}$ is the retarded time. In this
form, the vector potential can be analytically continued to complex
values of $O$, and then the projection onto real electromagnetic
fields as in section \ref{sec:Electromagnetism-in-complex} can be
performed. 

\noindent %

This model for spin is still too simple to lead to a description for
a quantum electron as a Kerr-Newman particle. An explanation for the
quantum behavior of spin is needed in this context. When one tries
to make sense of the Dirac equation as a quantum wave equation, one
is confronted with the zitterbewegung phenomenon. The Foldy\textendash Wouthuysen
transformation removes the zitterbewegung from the motion for positive
energy states and produces a non-radiating current density \cite{davidson_quantum_2007}.
One would think that a description of quantum fermions in the current
framework should incorporate some explanation for zitterbewegung for
Kerr-Newman particles. One such is given in \cite{davidson_lorentz-dirac_2012}
which is based on the classical Lorentz-Dirac equation embedded in
complex space-time. Another model for zitterbwegung has been suggested
by Burinskii, based on his gravitating bag model \cite{burinskii_gravitating_2015,burinskii_emergence_2015}.
The bag structure gives internal structure to the particle, and it's
no longer a simple point in complex space-time, but rather a closed
loop string in space. This ring acts as a waveguide for massless particle
excitations travelling the circumference of the ring, and this is
the model for zitterbewegung \cite{burinskii_kerrnewman_2014}. It
also makes a connection with string theory which is an interesting
way to obtain a connection with the Standard model of particle physics.
It seems reasonable to try and couple the Bohmian picture with one
of these zitterbewegung models. It's not obvious to me how to proceed
though. So this task remains as a topic for future research.

\section{The non-relativistic Schrödinger equation}

\noindent It is straightforward to parallel the treatment provided
here for the relativistic case for the the standard non-relativistic
Schrödinger equation. For the Gaussian case we can use Holland's results
\prettyref{eq:Holland trajectory} from \cite{holland_quantum_1995}.
We again embed these trajectories in complex space-time. In the non-relativistic
case, the trajectories are functions of the usual time variable $t$
rather than the world time $s$. Everything proceeds more or less
as in the relativistic case. The retarded times are still multivalued,
and the GAN continuation still generates all of the trajectories from
a single trajectory. The non-radiating property holds as well for
Schrödinger current \cite{davidson_quantum_2007}. 

\section{\label{sec:Interpretation-of-complex}Interpretation of complex space-time}

A reasonable question to ask is ``Don't we live in a real space-time
universe?'', and consequently ``doesn't that rule out most of this
paper?''. The simplest answer I can give is - maybe we don't live
in a real Minkowski space. Maybe we actually do live in complex space-time,
i.e. a 4 dimensional complex manifold endowed with a Minkowski metric
tensor, at least locally. Imagine that the universe is built up from
point particles moving around but always very near to the real hyperspace,
moving say along Bohmian trajectories and in some interaction with
one another. Then all of our experimental apparatuses, being made
up of such particles, are aware only of macroscopic distances in the
real space-time which is given empirical prominence because all the
matter is located there and large apparatuses are always therefore
embedded in this 4-dimensional real subspace. For large objects, and
for charged macro particles involving many elementary particles, I
expect everything to look like ordinary classical electromagnetism
in real space-time. All of the Bohmian trajectories for many particles
will result in much averaging, and only the dominant retarded time
for an extended macro particle will be found to contribute I suspect
(or more honestly hope). When dealing with a single small elementary
particle over small distances, I expect the corrections due to the
complex space-time effects to become visible. What we call quantum
mechanics might, in this fanciful picture, be the appearance of complex
space time effects for microscopic objects. The fuzziness of quantum
mechanics might be due to the extended particle nature of the electromagnetic
fields produced, and to the multi-valued nature of trajectories and
fields when small particles are being focused on. So, we think we
live in a quantum world of 4 real dimensions, and our classical world
is the large particle limit of this which obeys classical equations
of motion, but maybe in reality we are living in a multi-valued but
classical universe which consists of 4 complex dimensions, with the
matter concentrated on the real hyperspace, and what we call quantum
behavior, that we observe for small particles, is actually the effects
of the complex space-time modifying the Liénard-Wiechert potentials
and generating multi-valued fields and positions for small particles.
I grant that this is a speculative and fanciful picture, but in fact
it is in my mind when I think about these things. 

There is another possibility though, and this is slightly different,
but I think probably closer to what might be preferred by relativists.
Remember that there is always a projection to real fields on a real
hyperspace at the end of a calculation of the electromagnetic field
produced by some complex trajectory (or real trajectory embedded complex
space-time). So we always end up with real electromagnetic fields.
We can simply assume that classical electromagnetism should include
all of the roots that would be obtained by complex analytic continuation
first with the resulting fields then projected down on the real hyperspace.
This would then yield the same result as above where we treated the
space as actually a complex manifold. This would still yield multi-valued
fields and positions for particles. 

This is the sense that I meant when I said in the introduction that
I'm agnostic about whether the complex space-time was ontologically
real or not.

\section{An analytic universe interpretation of quantum mechanics }

\noindent Maybe quantum mechanics is a description of a multi-Riemann-sheeted
universe. Quantum superposition might be related to the GAN continuation
which is a form of linear superposition of different Riemann sheets.
It would need to be applied to the wave function itself, rather than
to the trajecories, and so it might come about in a second quantization
of the wave function treated as a dynamical field. Quantum mechanics
is perfectly linear, as measured in many experiments \cite{leggett_testing_2002,arndt_testing_2014,bialynicki-birula_linearity_2005}.
This is surprising. If quantum mechanics has some deep dynamical origin,
how does it arise that linear superposition of wave functions is exactly
correct? In the classical world linearity is always only approximate.
The generalized analytic continuation based on a partition of unity
argument gives exact linearity with absolutely no non-linear term.
And yet, it arises from the mathematics of general relativity, electromagnetism,
and Bohmian mechanics in complex space-time. One can start with the
many worlds or Everett interpretation of quantum mechanics \cite{everett_theory_1973},
and ask if the many worlds can be replaced by a many-Riemann-sheeted
interpretation of quantum mechanics? Regarding the Einstein-Rosen
bridge and entanglement ideas \cite{maldacena_cool_2013,susskind_dear_2017},
perhaps these too can be interpreted in terms of Riemann sheets. One
might imagine that collapse of the wave function is related to an
abrupt change in the weighting factors for different Riemann sheets
as in the generalized analytic continuation idea, or perhaps in a
change in the Riemann sheet structure itself. This is because a wave
function collapse would cause a change in the Bohmian trajectories
and their weighting functions and perhaps in their analytic Riemann
sheet topology. The results of this paper together with \cite{davidson_lorentz-dirac_2012}
and the complex manifold techniques of general relativity seem to
be suggesting such an analytic universe interpretation as a possibility. 

\section{Conclusion}

\noindent If Kerr-Newman particles are coupled to the electromagnetic
fields in complex space-time, and are moving along Bohmian trajectories,
then the trajectory equations become multi-valued with two or more
Riemann sheets. If one accepts the generalized analytic continuation
(GAN) proposal as valid, then for the Gaussian wave functions studied
here, all the Bohmian trajectories can be continued into one another,
except for a set of measure zero. This explains how such Kerr-Newman
particles need not radiate electromagnetically. It also allows the
point particle to act as an extended charge distribution or jellium
which would generally have a finite electrostatic energy. The hydrodynamic
or jellium model makes an experimentally testable prediction involving
bremsstrahlung as well \cite{davidson_predictions_2004}. In the standard
Bohmian interpretation, one is not allowed to couple the charged particles
in this way. So these results are an application of Bohmian mechanics
to a hybrid interpretation combining the Kerr-Newman complex space-time
techniques with Bohmian mechanics. The results here in no way detract
from the standard Bohmian interpretation, but add a possible new application
for Bohmian mechanics.

Although metric nonlinearities are ignored in this paper, the complex
space-time embedding is not ignored, and the motivation for this embedding
comes from general relativity. Therefore, the multi-sheeted and multi-valued
trajectories and resulting fields are also a direct consequence of
general relativity that cannot be ignored, even in the very weak gravity
limit. Thus we have a flat space theory in which general relativity
is playing a very major role, and in fact it's offering potential
explanations for some of the deepest mysteries of quantum mechanics.
It also leads to a way to avoid the infinite self-energy of a static
point particle in classical electromagnetism by making a point charge
behave as an extended particle.

For multi-particle systems the multiplicity of Riemann sheets increases
exponentially. These results beg the question - Can an interpretation
of quantum mechanics be developed that is based on this multi-sheeted
structure, in a sense mapping the spirit of the many-worlds or Everett
interpretation \cite{everett_theory_1973} into a single analytic
complex world manifold with many Riemann sheets? 
\begin{acknowledgements}
I would like to thank the two reviewers for this paper who provided
valuable insights and suggestions. I would also like to thank Ezra
Newman for his helpful correspondence.
\end{acknowledgements}

\noindent 

\noindent \bibliographystyle{plain}
\bibliography{FOP_Merged_Total}

\end{document}